# Device and Circuit Interaction Analysis of Stochastic Behaviors in Cross-Point RRAM Arrays

Haitong Li, *Student Member, IEEE*, Peng Huang, Bin Gao, *Member, IEEE*, Xiaoyan Liu, *Member, IEEE*, Jinfeng Kang, *Senior Member, IEEE*, and H.-S. Philip Wong, *Fellow, IEEE*

***Abstract*—Stochastic behaviors of resistive random access memory (RRAM) play an important role in the design of cross-point memory arrays. A Monte Carlo compact model of oxide RRAM is developed and calibrated with experiments on various device stack configurations. With Monte Carlo SPICE simulations, we show that an increase in array size and interconnect wire resistance will statistically deteriorate write functionality. Write failure probability (WFP) has an exponential dependency on device uniformity and supply voltage ($V_{DD}$), and the array bias scheme is a key knob. Lowering array $V_{DD}$ leads to higher effective energy consumption (EEC) due to the increase in WFP when the variation statistics are included in the analysis. Random-access simulations indicate that data sparsity statistically benefits write functionality and energy consumption. Finally, we show that a pseudo-sub-array topology with uniformly distributed pre-forming cells in the pristine high resistance state is able to reduce both WFP and EEC, enabling higher net capacity for memory circuits due to improved variation tolerance.**

***Index Terms*—Resistive random access memory (RRAM, ReRAM), non-volatile memory, variability, statistical analysis, Monte Carlo, compact model, cross-point array, SPICE.**

## I. Introduction

RESISTIVE random access memory (RRAM) is a promising building block of nonvolatile information storage systems for data-centric applications [1]-[4], leveraging its simple structure, high performance, and good scalability [5]-[12]. Variability of RRAM characteristics imposes constraints on memory cell/array design, but also inspires new applications such as stochastic neuromorphic systems and physical unclonable function [13]-[15]. Hence, a deeper understanding of the device and circuit variability is essential to architect RRAM for memory and logic applications. Previous research efforts were mainly focused on the intrinsic device-level variability and corresponding physical

Manuscript received 2017. This work was supported in part by the Member Companies through the Stanford Non-Volatile Memory Technology Research Initiative (NMTRI), STARnet SONIC, the NCN-NEEDS Program within the National Science Foundation under Contract 1227020-EEC, and by Semiconductor Research Corporation, and in part by the National Natural Science Foundation of China under Grant 61421005 and Grant 61334007.

H. Li and H.-S. P. Wong are with Department of Electrical Engineering and Stanford SystemX Alliance, Stanford University, Stanford, CA 94305, USA (haitongl@stanford.edu, hspwong@stanford.edu).

P. Huang, X. Liu, and J. Kang are with Institute of Microelectronics, Peking University, Beijing 100871, China (kangjf@pku.edu.cn).

B. Gao is with Institute of Microelectronics, Tsinghua University, Beijing 100871, China.

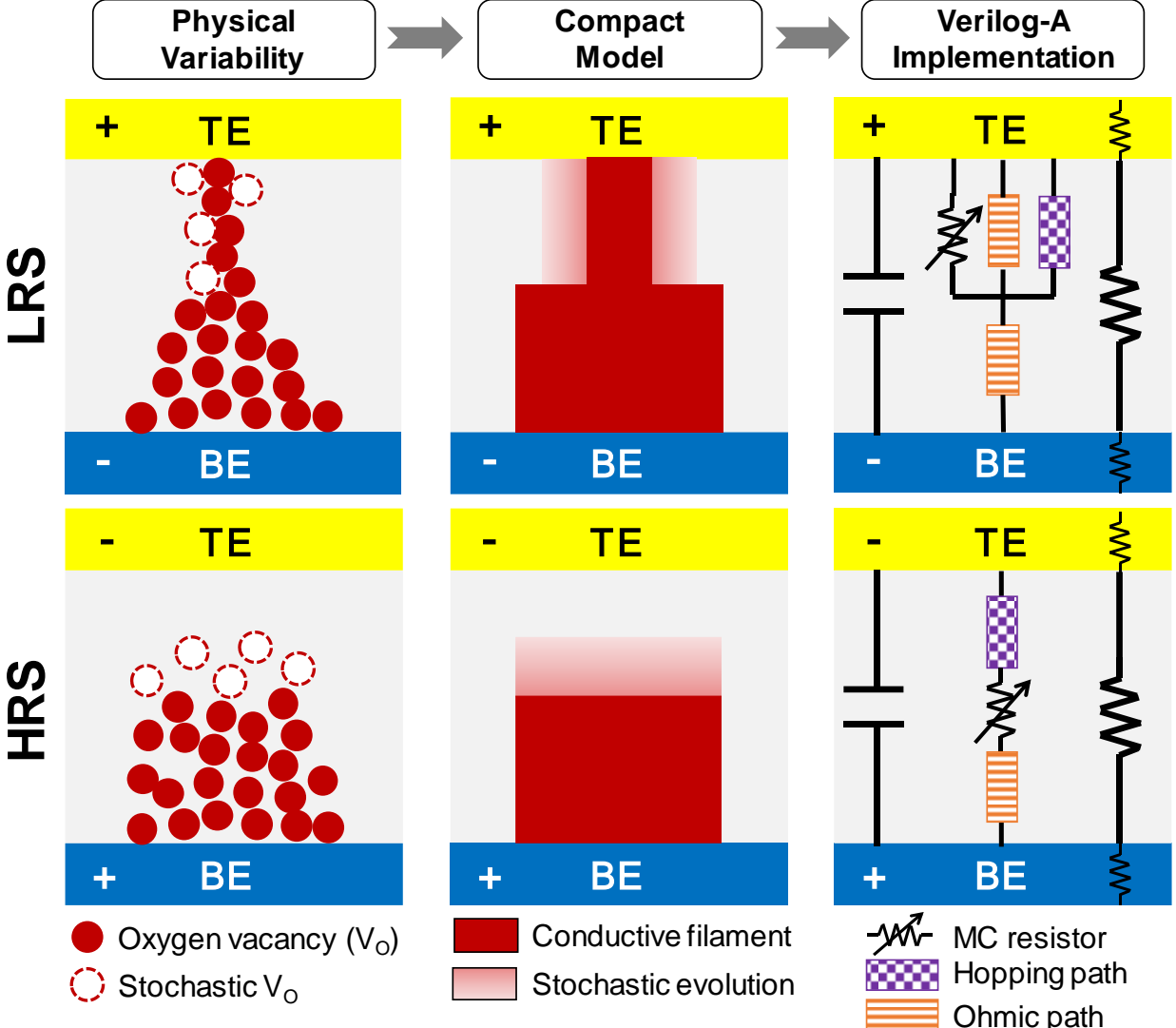

Fig. 1. Monte Carlo (MC) RRAM model hierarchy: physical variability, compact model, and Verilog-A implementation.

mechanisms, which were correlated with the choice for the materials, stack configuration, as well as operation conditions [16]-[21]. Several array-level studies have analyzed the impact of typical device characteristics (without variability) on array write/read operations [22]-[27], while others further investigate the variability impact on read performance using simplified resistor models [28], [29]. These studies either used over-simplified device models that were not able to capture essential device behaviors, or failed to incorporate cell variations into circuit-level analysis. Thus, a cross-layer analysis that links the physical picture of intrinsic device variability with circuit-level stochastic behaviors is still lacking. In this work, we scrutinize the stochastic behaviors of cross-point RRAM arrays from the perspective of device and circuit interaction. Large-scale SPICE simulations are performed using a Monte Carlo (MC) compact model of RRAM, which is built upon a stochastic conductive filament (CF) evolution model and calibrated by device measurements. Through a suite of variation-aware circuit analysis, we probe into the array write functionality, reliability, energy, and random-access behaviors in a statistical manner. The 'translation' of variations from device level into circuit level is neither linear additive nor analytical, and larger arrays tend to 'amplify' the circuit variations translated from device tails, especially for low-power embedded applications. This key observation renders the methodology in this work essential for variation-aware design and optimization of RRAM arrays.

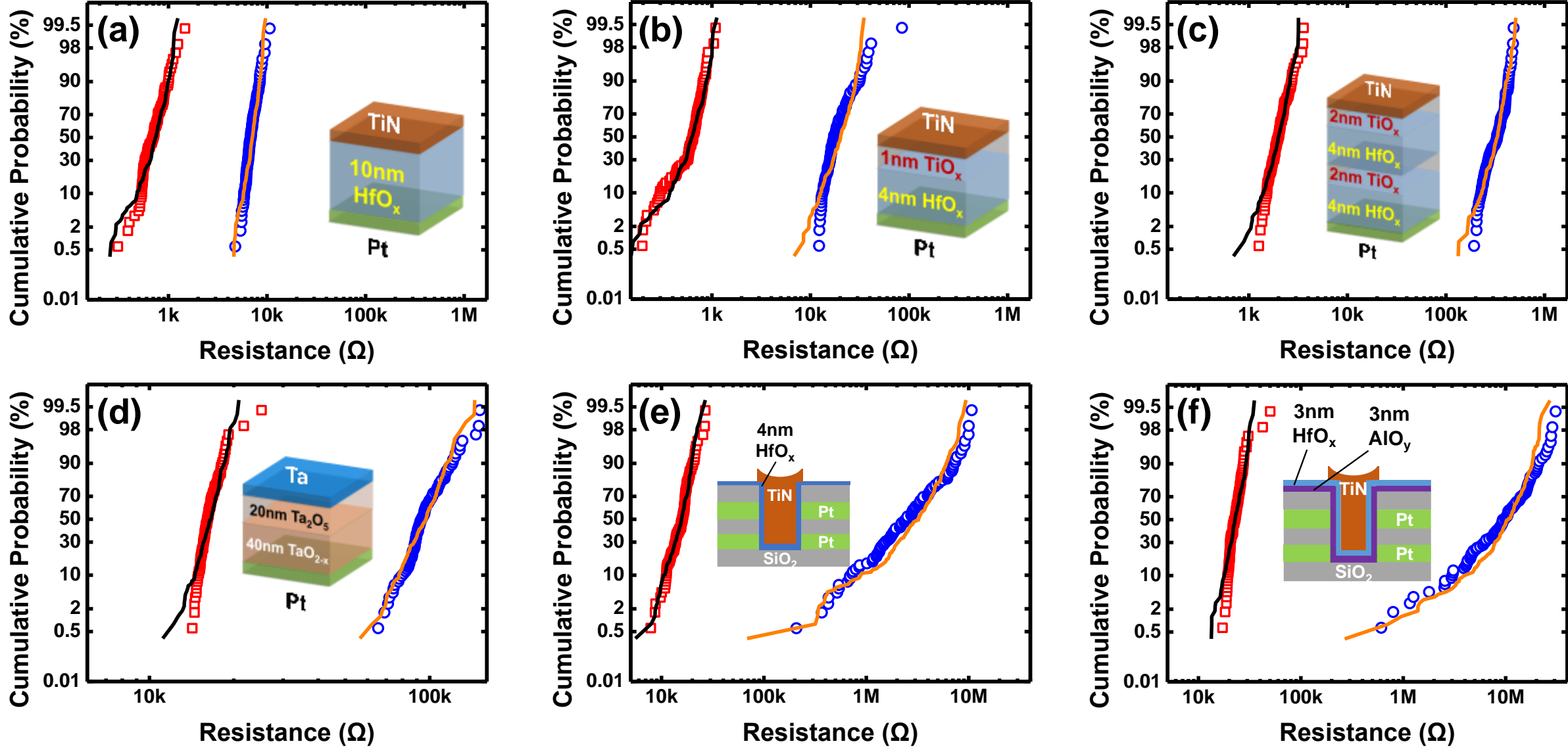

Fig. 2. Measured (symbols) and modeled (lines) statistical distributions of $R_{HRS}$ and $R_{LRS}$ on a variety of oxide-based RRAM devices, including (a) $HfO_x$ RRAM, (b) $HfO_x/TiO_x$ bi-layer RRAM, (c) $HfO_x/TiO_x/HfO_x/TiO_x$ multi-layer RRAM, (d) $Ta_2O_5/TaO_{2-x}$ RRAM, (e) $HfO_x$ 3D vertical RRAM, and (f) $HfO_x/AlO_y$ 3D vertical RRAM. Each inset illustrates the material stack configuration for the measured samples.

## II. Monte Carlo Compact Model

The variability of CF geometry leads to the widely observed random variations of high resistance states (HRS, bit '0') and low resistance states (LRS, bit '1') [16]. Specifically, it has been shown that the resistance distributions of LRS ($R_{LRS}$) and HRS ($R_{HRS}$) result from the fluctuations in the CF radius and the tunneling gap distance, respectively [16]-[18]. A physics-based compact model that captures device variability is required to investigate the circuit-level stochastic behaviors. Developing a method to efficiently incorporate the variations in a compact model is even more critical given the complexity of some of the in-memory computing circuits being designed [30]-[32]. Fig. 1 shows the model hierarchy, from physics of variability to Verilog-A sub-circuit implementation of the compact model. Physical variability is described by the stochastic location of oxygen vacancies around the CF and the gap region. The variations of electrical characteristics of RRAM are then described by the intrinsic variability of CF geometry in terms of length and width. In the Verilog-A implementation, the median CF evolution is modeled by the hopping/ohmic paths as a deterministic sub-circuit [33], while the stochastic CF geometry is modeled by a Monte Carlo (MC) resistor as a variation sub-circuit. For LRS, the MC resistor and deterministic CF part are in parallel, modeling the CF width variations. For HRS, the MC resistor is in series with the deterministic CF part, modeling the CF length or gap distance variations. The dominant equations that describe the deterministic CF evolution behaviors follow the same set as those in an experimentally calibrated compact model [33]-[35], and the variation sub-circuit is calibrated using the statistical data from measurements. The MC compact model captures the essential variations in an efficient manner, while retaining the physics and accuracy of the deterministic part. The generality of MC variation sub-circuit representation is confirmed by the measurements on a variety of RRAM devices, as shown in Fig. 2. The measured devices include (a) $HfO_x$ RRAM, (b) $HfO_x/TiO_x$ bi-layer RRAM, (c) $HfO_x/TiO_x/HfO_x/TiO_x$ multi-layer RRAM [36], (d) $Ta_2O_5/TaO_{2-x}$ RRAM, (e) $HfO_x$ 3D vertical RRAM [37], and (f) $HfO_x/AlO_y$ 3D RRAM [38]. The fitting procedure for each type of device is as follows: statistical distributions of $R_{LRS}$ and $R_{HRS}$ are first obtained by 100-cycle SET/RESET measurements, with median and standard deviation (SD) extracted assuming Gaussian distributions. The deterministic sub-circuit in the compact model captures the median values of measured $R_{LRS}$ and $R_{HRS}$ following the principle in [34]. To add the variations, the variation sub-circuit, i.e. the MC resistor, takes the extracted SD values as an input parameter, and generates normal distributions through statistical MC simulations in HSPICE [39]. Note that the distributions in (a)–(f) present different curvatures and tails, in both experimental and modeled data. For experiments, the statistical distribution results from the intrinsic randomness of CF position, size, and even morphology. The measured distributions may keep varying with more cycles involved (if insufficient samples are measured initially, or if some irreversible changes in the RRAM occur, such as endurance cycling degradation). For modeling, HSPICE uses non-identical random seeds to generate specified distributions from device to device, which mirrors such dynamics of tails and curvatures observed in our measurements. This model formulation enables the compact model to keep its analytical nature for modeling both deterministic and stochastic behaviors efficiently, which further enables us to conduct the array-level circuit/device interaction analysis. A case study using the device characteristics in Fig. 2(c) is performed and discussed in the following sections.

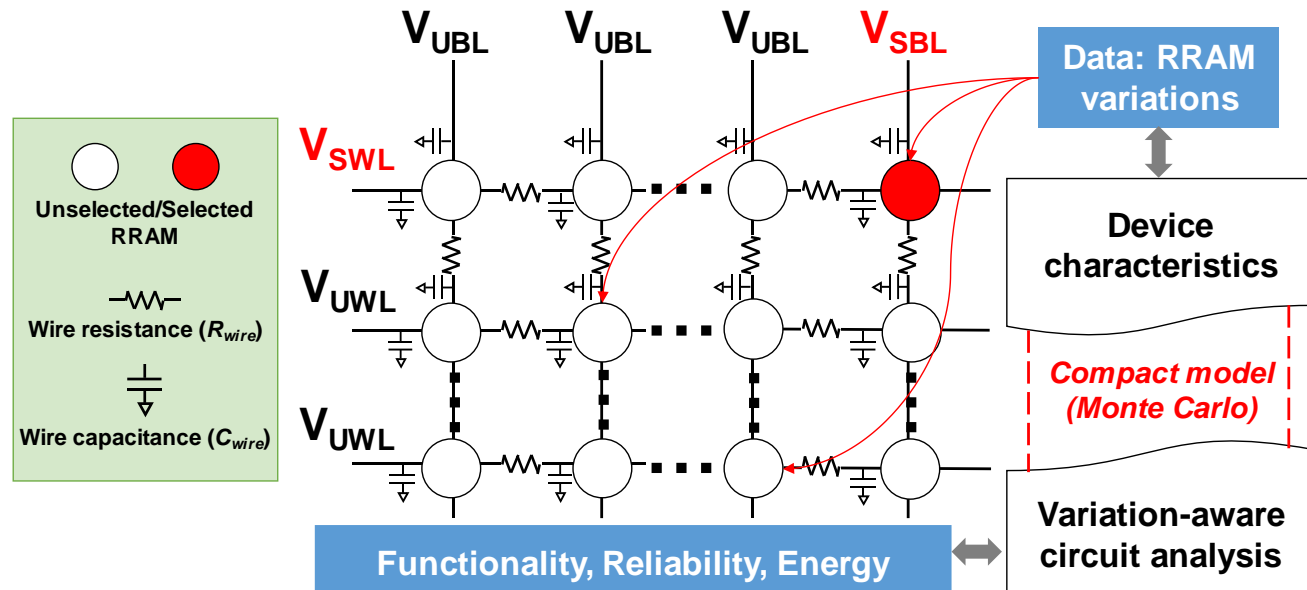

Fig. 3. Schematic of a cross-point RRAM array and the corresponding framework for device and circuit interaction analysis (SWL: selected word line; UWL: unselected word line; SBL: selected bit line; UBL: unselected bit line; WAV: write access voltage; WFP: write failure probability).

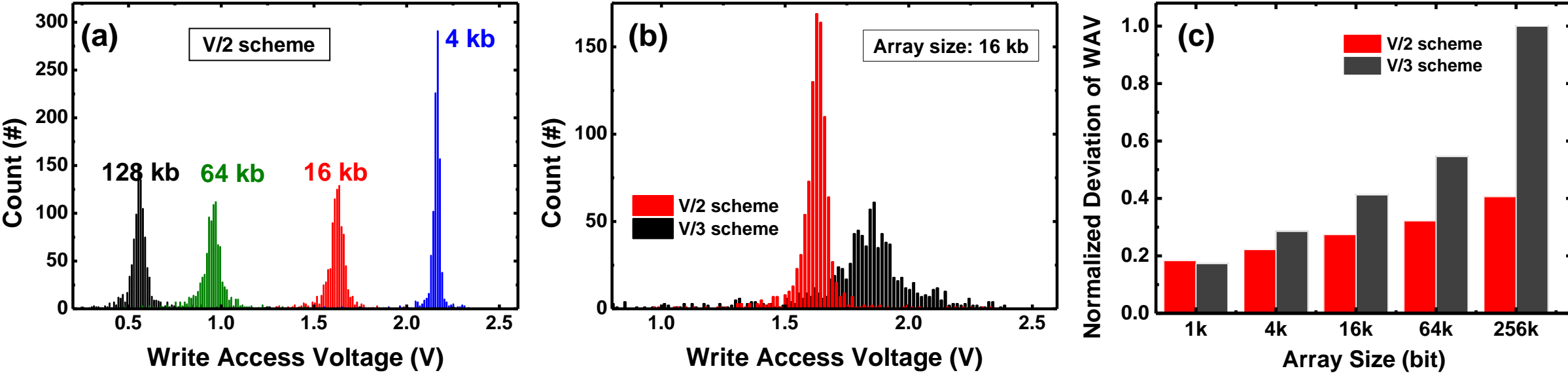

Fig. 4 Variation-aware write functionality assessment of n×n cross-point RRAM arrays. (a) Statistical distributions of write access voltage (WAV) for the worst-case selected cell with various array sizes. The median WAV decreases and the distribution diverges as array size increases. Each spread is generated from 1000 cycles. (b) Statistical distributions of WAV under different bias schemes. V/3 scheme provides higher median WAV as well as higher variance, whereas V/2 scheme leads to a tighter distribution. (c) Comparison of WAV variance between V/2 and V/3 schemes as a function of array size. The variance induced by V/3 scheme increases rapidly with larger arrays.

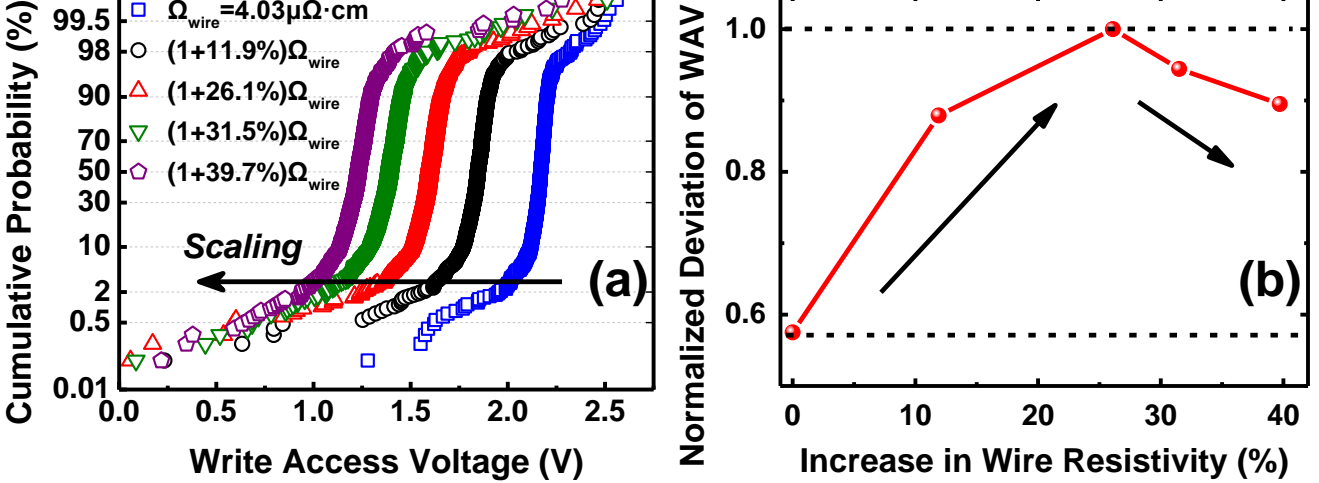

Fig. 5 Impact of interconnect wire resistance on the WAV distributions. (a) Statistical WAV distributions of a 4-kb array as wire resistance scales. Interconnect scaling will push the distributions towards a lower margin. (b) WAV variance as a function of the increase in wire resistivity due to scaling.

## III. Stochastic Behaviors in RRAM Arrays

To illustrate how RRAM variations are 'translated' into the circuit-level stochastic behaviors, n×n cross-point array structure without selection devices is used as an example. Fig. 3 illustrates the analysis framework incorporating the measured/modeled RRAM variations. In the following simulations, the technology node is assumed to be 22 nm for cross-point arrays. Metal wires have a 44-nm pitch, an aspect ratio (AR) of ~2, and a sheet resistance of 1.405 Ω/square. The capacitance of the wires is 1.045 fF/μm. Therefore, in the array model, wire resistance ($R_{wire}$) is 2.81 Ω/cell and wire capacitance ($C_{wire}$) is 0.046 fF/cell [40]. RRAM MC model is calibrated to experimental measurements of the device in Fig. 2(c). A worst-case scenario is adopted, where the farthest-corner selected cell is in HRS and all the unselected cells are in LRS [22]. V/2 scheme and V/3 scheme, the two major bias schemes for cross-point RRAM arrays, are used for write operations:

$$V/2: \quad V_{SWL} = V_{DD},\ V_{SBL} = 0,\ V_{UWL} = V_{UBL} = V_{DD}/2 \tag{1}$$

$$V/3: \quad V_{SWL} = V_{DD},\ V_{SBL} = 0,\ V_{UWL} = V_{DD}/3,\ V_{UBL} = 2V_{DD}/3. \tag{2}$$

With 1000-cycle MC simulations in HSPICE for each device/circuit configuration, three aspects of stochastic behaviors in cross-point RRAM arrays are investigated: write functionality, reliability, and energy consumption. Random-access simulations are conducted additionally to analyze write operations on 64-kb arrays with D2D variations, which probe into the array stochastic behaviors at a higher level.

### A. Write Functionality

Write access voltage (WAV) is defined as the actual voltage drop on the selected cell in a memory array, which is lower than the array voltage supply $V_{DD}$ due to interconnect IR drop [35], [41]. Fig. 4 shows the variation-aware assessment of cross-point RRAM arrays. Statistical distributions of WAV for various array sizes are obtained by simulating write operations under the aforementioned V/2 scheme ($V_{DD}$ = 2.6 V), as shown in Fig. 4(a). The WAV distributions result directly from the device-level resistance variations. Correspondingly, the difference in specific resistance patterns due to device resistance variations are 'translated' to differences in sneak path configurations and IR drop along interconnect wires. Thereby, WAV diverges from its nominal value. Fig. 4(a) shows that WAV for 4-kb array has a tight distribution and a higher median value, whereas the distribution tends to spread out in larger arrays. This is mainly because the possible number of configurations of either sneak paths or data patterns grows exponentially with array size. This observation, on the other hand, implies that the 'translation' of variations from device measurables (eg., $R_{LRS}$ and $R_{HRS}$) to circuit outputs (eg., WAV) is nonlinear. Therefore, MC simulation framework is necessary to help visualize and quantify such translation and interaction. Fig. 4(b) shows that using V/3 scheme results in relatively higher median WAV but worse distributions compared with V/2 scheme. This illustrates the dominated role of unselected cells (cross-points of all the unselected WLs and BLs) in WAV distributions for the in the V/3 scheme write operation. In V/3 scheme, all the unselected cells have biases near $V_{DD}$/3. In comparison, the unselected cells in V/2 scheme have near-zero biases, which reduces the impact of stochastic configurations of sneak paths on WAV distributions. Such difference between V/3 and V/2 schemes becomes more significant in larger arrays, as indicated by Fig. 4(c). In addition to the scaling up effect of array sizes, the impact of interconnect scaling down is quantified as well. Fig. 5(a) shows the statistical distributions of WAV as interconnect wire resistance increases due to scaling-induced scattering and skin effects [40], [41]. The median shift with the increase in wire resistivity is attributed to the degradation of IR drop along the selected signal path. SD of the WAV is extracted and normalized, as shown in Fig. 5(b). The largest $R_{wire}$ does not necessary correspond to the widest distributions. This trend is in line with the median's shift towards lower write margin shown in Fig. 5(a). Specifically, the variations of RRAM cells become less important as $R_{wire}$ increases with larger IR drop, and thus the WAV distributions tend to become tighter (as $R_{wire}$ variation is not included in the model).

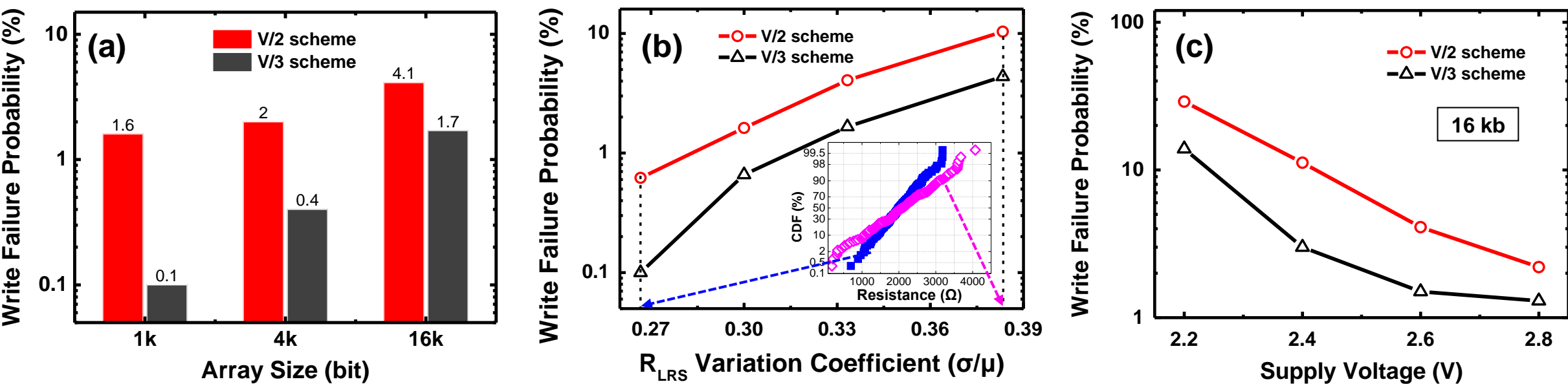

Fig. 6 Quantitative analysis of write failure behaviors in cross-point RRAM arrays. (a) Write failure probability (WFP) as a function of array size and bias scheme. Each probability is obtained from 1000-cycle operations. (b) Impact of device uniformity on WFP of a 16-kb array under V/2 and V/3 schemes. Inset illustrates how the variation coefficients correspond to $R_{LRS}$ distributions. (c) WFP of a 16-kb array increases exponentially under V/2 and V/3 schemes as array $V_{DD}$ decreases.

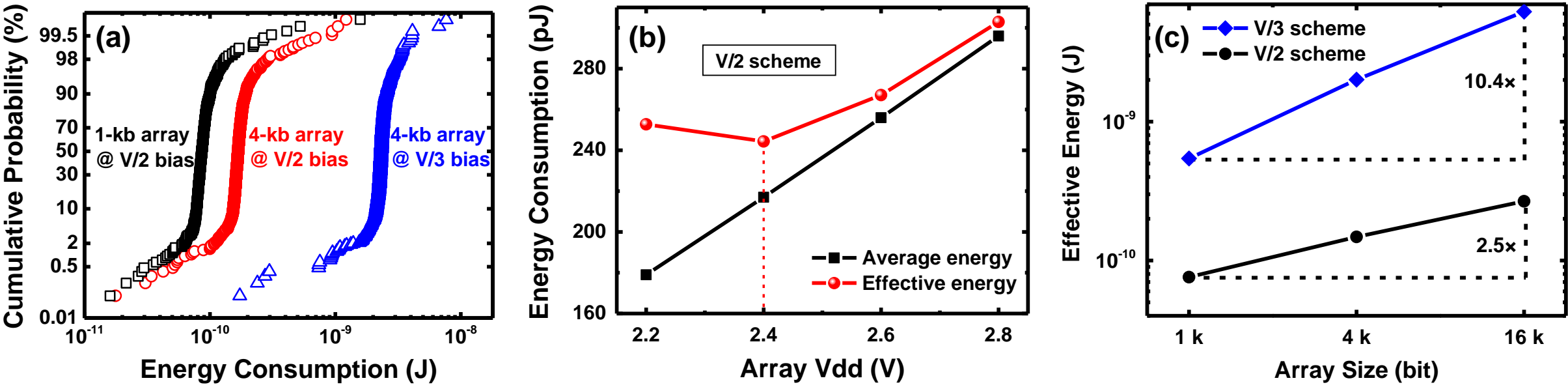

Fig. 7 Array energy consumption from a statistical perspective. (a) Statistical distributions of total energy consumption during array write operations. Each spread contains data from 1000 cycles. (b) Comparison of the average energy and effective energy per cycle as a function of array $V_{DD}$ under V/2 scheme. The effective energy consumption reversely increases under low array $V_{DD}$. (c) Effective energy consumption as a function of array size and bias scheme. V/2 scheme leads to lower effective energy consumption compared with V/3 scheme.

### *B. Write Reliability*

Besides the write functionality discussed above, array reliability in terms of statistical write failure behaviors is also highly dependent on the interaction between device variations and circuit conditions. Write failure probability (WFP) is quantified by simulating SET operations to analyze such dependency in cross-point RRAM arrays. For other data patterns, WFP analysis of RESET operations follows the same principles as SET operations. A write failure event occurs when the access voltage is not sufficient to switch (i.e., SET/RESET) the selected RRAM cell. And WFP is defined as the ratio of failed write events to total write events. As shown in Fig. 6(a), the WFP induced by device-level variations generally increases with larger array sizes. V/3 scheme has lower WFP compared with V/2 scheme. This is mainly due to higher median WAV in the V/3 scheme even though the V/3 scheme has relatively higher degree of variance. Such higher variance also leads to a rapid increase in WFP as array size scales up. High median voltage stress in V/3 scheme might cause earlier endurance failure [42], whereas the low-median low-variance V/2 scheme could lead to better endurance performance. There are tradeoffs between different reliability metrics such as WFP and endurance. Fig. 6(b) shows the impact of device uniformity on WFP, where the median is fixed and the variation coefficient (VC) (ratio of SD to mean, σ/μ) of $R_{LRS}$ is varied for a sensitivity analysis. It is shown that a 44% increase in VC of $R_{LRS}$ distributions leads to 16× and 43× increase in WFP under V/2 scheme and V/3 scheme, respectively. Hence, cross-point arrays biased using the V/3 scheme is more sensitive to RRAM variations as compared to the V/2 scheme. Under both bias schemes, a linear VC improvement in resistance uniformity will be able to provide an exponential decrease in WFP, suggesting that there is a broad optimization space for device engineering [36], [43], [44]. In addition, WFP is closely related to the supply voltage for RRAM arrays. As shown in Fig. 6(c), linearly lowering the supply voltage causes an exponential increase in WFP for both bias schemes. A 20% decrease in array $V_{DD}$ results in 12× and 10× increase in WFP under V/2 scheme and V/3 scheme, respectively. Higher $V_{DD}$ lowers WFP but may raise other reliability concerns such as endurance and write disturbance on half-selected cells in cross-point arrays [45]. These results and tradeoffs imply the importance of adaptive write circuitry and error correction. It is worth noting that the absolute values of WFP do not exactly correspond to the bit error rates in real memory systems. This is merely because bit error rate (with a unit of ppm/ppb) is based on $>10^6$ monitored events whereas WFP (with a unit of %) is based on full-circuit MC simulations with $10^3$ samples. Therefore, given specific device characteristics, the circuit conditions (e.g., the supply voltage) are chosen to quantify WFP at a level

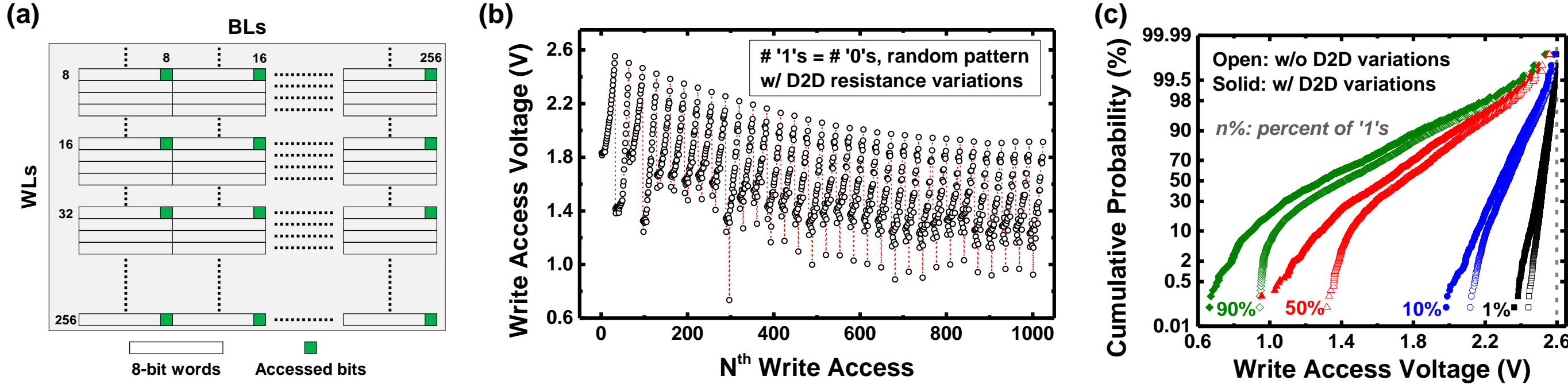

Fig. 8 Random access simulations at the array level enabled by the MC RRAM model. (a) In a 64-kb array, the last bits of the 8-b words (every 8th row) are accessed during simulations (1024 writes in total). (b) WAVs on the accessed cells following an address sequence of (8*x*, 8*y*) where *x* is from 1 to 32 (WL select) and *y* is from 1 to 32 (BL select). Bit '1's and '0's are randomly and equally mapped within the 64-kb array, with device-to-device (D2D) resistance value variations included. (c) Statistical distributions of WAV in the 64-kb array with various levels of data sparsity (percentage of '1's) initialized.

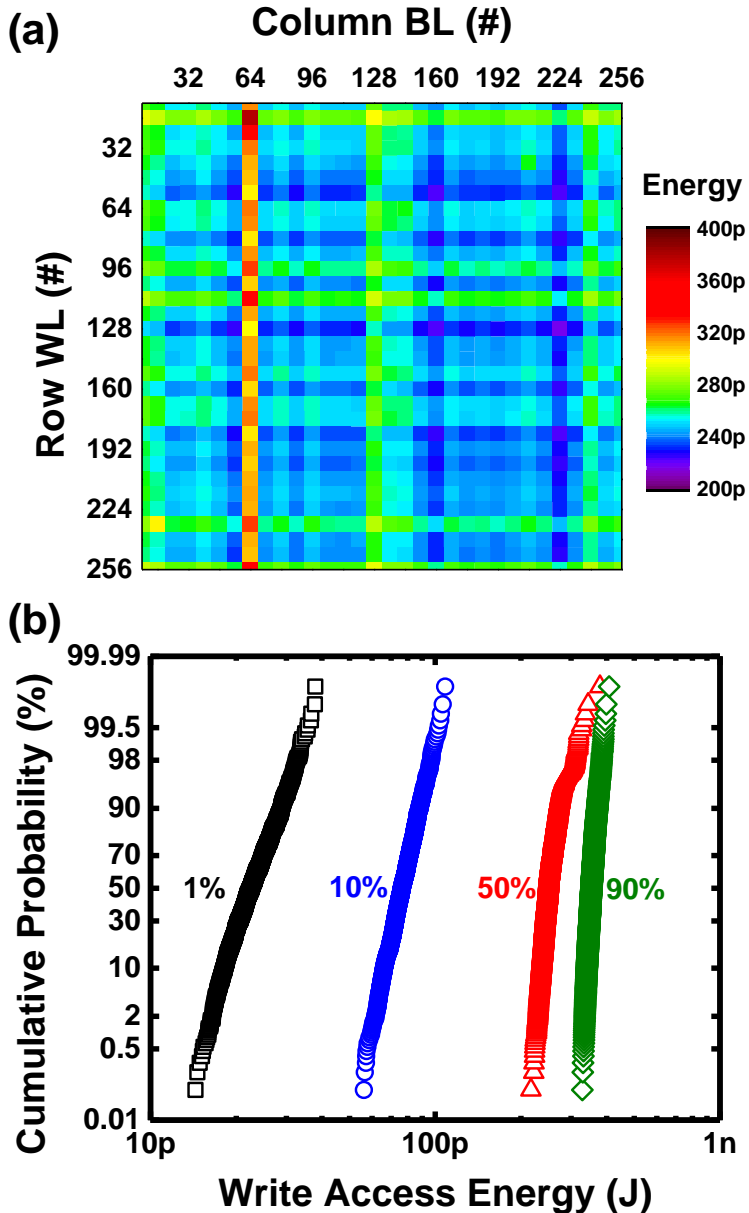

Fig. 9 (a) Simulated write energy map attained from a 64-kb array (1024 cells accessed in total). Bit '1's and '0's are randomly and equally distributed. (c) Statistical distributions of write energy consumption in the 64-kb array with different levels of data sparsity (percent of '1's) initialized.

commensurate with the size of the MC samples used. By linearly sampling $V_{DD}$, WFP can be monitored in a wide range and be further extrapolated to the real system specifications.

### *C. Energy Consumption*

Energy efficiency is one of the major design objectives for architecting non-volatile memory sub-systems [1]. A deeper analysis from the statistical perspective is essential but rarely provided. Using MC simulations in HSPICE, the total energy consumption during write operations is obtained and analyzed. As shown in Fig. 7(a), device variations also result in a significant spread of energy consumption at the memory array level, which can lead to substantial degradation in energy efficiency. For instance, in 1-kb arrays biased under V/2 scheme, a 38% deviation below nominal $R_{LRS}$ of unselected cells results in a 45× deviation above nominal array energy consumption. Compared with the V/2 scheme, the V/3 scheme shows an order of magnitude higher energy consumption on average, mainly due to more sneak paths among unselected cells. The leakage current through these unselected cells with near $V_{DD}/3$ bias contributes substantial static energy consumption. Since the major variation source for these MC simulations is the worst-case array resistance pattern with cell resistance variability, the tails in energy consumption distributions correspond to the outliers of static energy due to unselected cells with abnormally low resistances. A missing piece in the analysis above is the write failure statistics. Statistically speaking, the energy consumption during failed write events is wasted. WFP should be taken into account to provide a metric for an effective energy consumption (EEC), which can be defined as:

$$EEC = \frac{Average\ Eenergy}{1-WFP}\ (J\ /\ \mathrm{op})\ . \qquad (3)$$

Fig. 7(b) compares the average energy and EEC under V/2 scheme for a range of array $V_{DD}$. The average energy consumption decreases with $V_{DD}$ as expected, whereas the EEC first drops and then rises. This is attributed to the low-$V_{DD}$-induced write failure. EEC can be roughly thought of the actual 'work' required for each successful write operation. Even though the specific values of such 'work' may change if error correction, adaptive write, or other techniques are used in the peripheral circuits, the key point of the analysis here still holds: as long as there exist failure events induced by intrinsic variability, there will be extra energy overhead. Fig. 7(c) shows the comparison between V/2 scheme and V/3 scheme in terms of EEC. It is shown that V/2 scheme is more energy efficient with an order of magnitude lower ECC than V/3 scheme. This is a race condition where V/2 scheme has relatively higher WFP but much lower leakage-induced static energy. The net effect of such interaction leads to lower EEC under V/2 scheme, again indicating the predominant role of unselected cells in overall energy efficiency. As array size increases from 1 kb to 16 kb, the leakage paths under V/3 scheme result in even larger spread of energy consumption and the rapid increase in WFP, which is reflected by the >10× increase in EEC.

### *D. Random Access Statistics*

Capturing device variations efficiently in the MC RRAM model enables investigations of random-access operations at the array level, which is a critical step towards chip-level screening and analysis. As a case study, 1024 random-access write operations on a 64-kb array are simulated, where the last bits of the 8-b words are accessed every 8th row in the whole array, as illustrated by Fig. 8(a). First, bit '1's and '0's are randomly and equally initialized in the 64-kb array with resistance value variations incorporated by the model. Then, WAVs are simulated for the 1024 accessed cells, as recorded in Fig. 8(b). It can be observed that the WAV generally tends to decrease as the accessed location gets farther away from voltage supply, with fluctuations around a certain range of rows and columns that are being accessed. Such behaviors can be quite important for array-level write verification, yet they could not be captured without variation-aware modeling. More complex and realistic scenarios would have various levels of data sparsity (percentage of '1's). Corresponding analyses are performed by mapping different sparsity into the array data patterns and simulating cell-by-cell write operations. As shown in Fig. 8(c), D2D statistical distributions of WAV are attained over a wide range of array data sparsity. The WAV degrades less if when bit '0's are dominant in array data patterns, which is owing to the reduced leakage current and therefore wire IR drop in the crossbar arrays. It is shown that with D2D resistance variations considered, there is an additional decrease in WAV for all cases, and the difference becomes larger when bit '1's become dominant in array data patterns. Changing sparsity from 50% to 10% or 1%, the WAV is greatly improved. This indicates that data coding might be a useful or even necessary tool for RRAM system management. In addition to write functionality during random accesses across the 64-kb array, the energy consumption per access is attained for the equal '0'/'1' data pattern. Because of various sources of variations such as location and resistance, the array energy consumption per access depends not only on the accessed cell characteristics (e.g., resistance value), but also on the location of the cell and its spatial relationship with adjacent blocks. Therefore, obtaining the energy consumption map during the random accesses is the most direct way to visualize and analyze the array status. Fig. 9(a) shows such an energy consumption map with respect to cell locations during random-access write operations. Access-to-access difference of energy consumption can be observed from the plot. This is due to device-level variations and the resulting array-level, location-based variations. It can be seen that peak energy events occur along the 64th column of the 64-kb array. The specific array data pattern and the cell resistances play a joint role in such results. Additionally, impact of data sparsity has big impact on the array-level energy consumption. As shown in Fig. 9(b), operating the RRAM arrays with 1% data sparsity leads to >10× write energy reduction, statistically instead of ideally, compared with 50% '0'/'1' data patterns. This can be explained by the higher cell resistances and changes in sneak paths, which result in lower leakage-induced static energy.

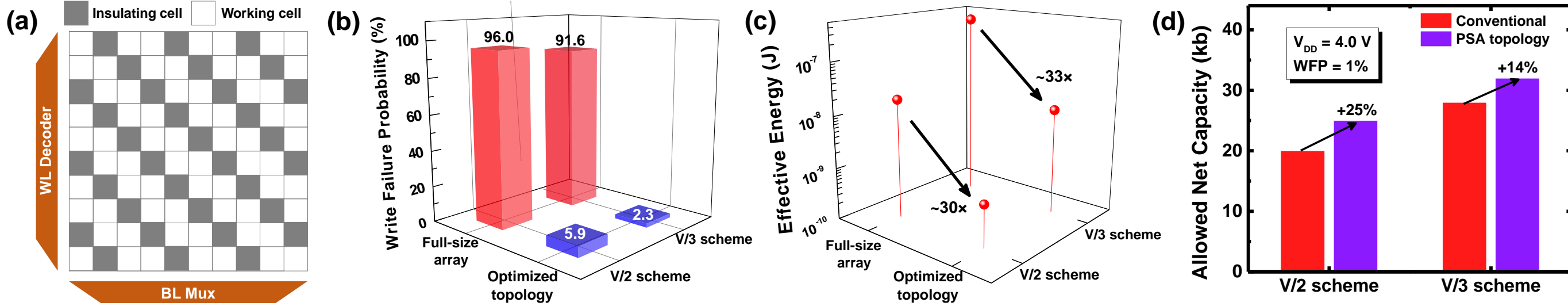

Fig. 10 Case study of a variation-tolerant pseudo-sub-array (PSA) topology for cross-point RRAM arrays. (a) Schematic of PSA topology where pre-forming cells are uniformly distributed. (b) Comparison of PSA topology and full-size cross-point array in terms of WFP (array size is 64 kb and $V_{DD}$ is 4.0 V). PSA topology significantly lowers the WFP by restraining array-level variations induced by sneak paths. (c) Comparison of PSA topology and full-size cross-point array in terms of EEC. PSA topology enables higher energy efficiency due to improved write reliability and reduction of array leakage current. (d) Allowed memory net capacity under conventional full-size and PSA topologies given the fixed 1% WFP.

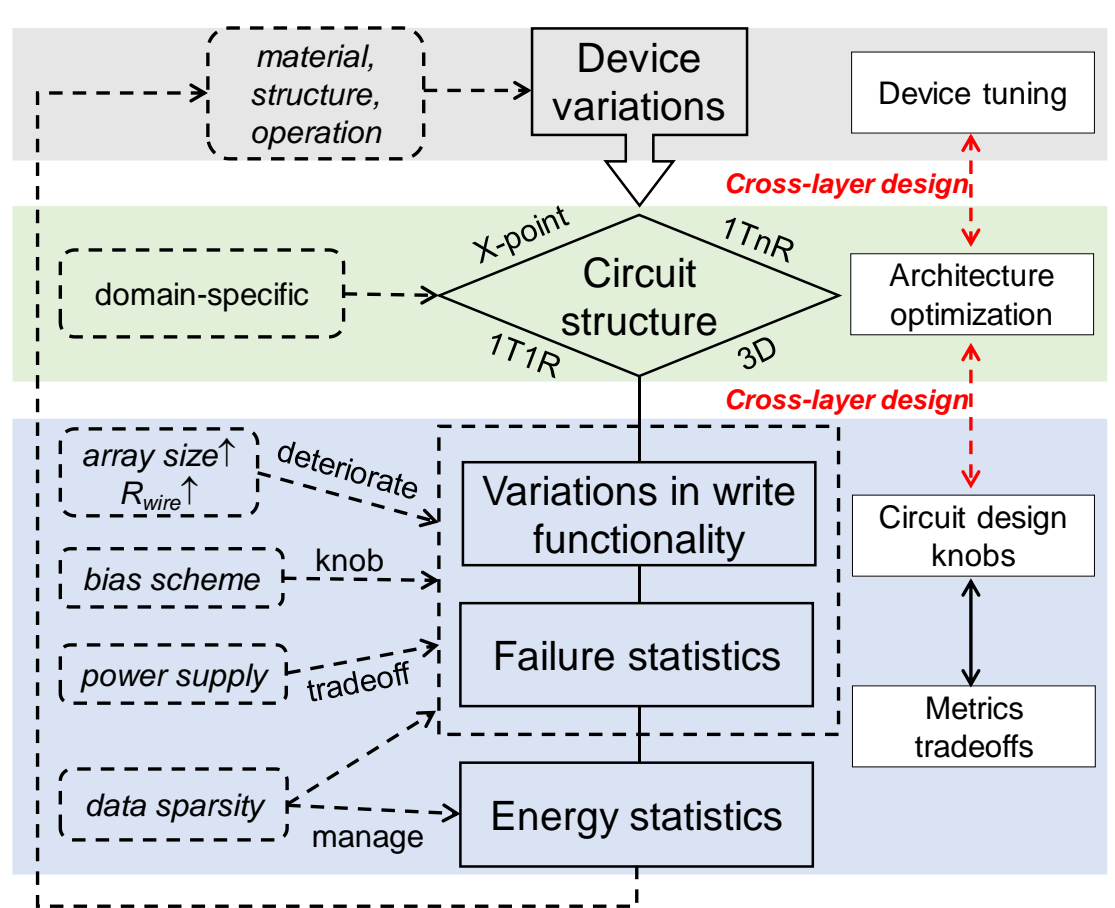

Fig. 11 Summary of design implications for RRAM memory arrays from a perspective of device and circuit interaction.

## IV. Array Optimization: Case Study & Guidelines

Based on the understanding of the stochastic behaviors in cross-point RRAM arrays, a case study for array optimization is performed. It was shown that uniformly distributed insulating (i.e., pre-forming high resistance) cells in RRAM arrays are able to help restrain sneak paths and thus reduce leakage current [45], [46]. Here, we consider such design's tolerance to variations. Fig. 10(a) shows the schematic of the pseudo-sub-array (PSA) topology, which is inspired by the partition of sub-arrays in a memory bank [47]. Since the working cell distribution is uniform in both WL and BL directions, memory sub-system design does not require fundamental changes except for the address encoding/decoding. Write operations on 64-kb arrays with 4.0-V $V_{DD}$ are simulated. As shown in Fig. 10(b), 4.0 V is insufficient for the write operations on full-size 64-kb arrays, with very high WFP. In contrast, the optimized PSA topology has significantly reduced WFP under the same $V_{DD}$ and bias schemes for 64-kb arrays. This is mainly because the uniform pattern of insulating cells effectively constrains the possible number of sneak path configurations, and thereby reduces the variations that are related to leakage current. Energy efficiency is also analyzed. As shown in Fig. 10(c), PSA topology reduces the EEC of 64-kb arrays by ~30× under V/2 scheme and ~33× under V/3 scheme. Improvement in EEC is a collaborative effect of lowered WFP and reduced static energy consumption. For cross-point RRAM arrays, the maximum allowed array size is usually limited by the write/read margin and write reliability of accessing individual cells [22]. A large array means large IR drop and reduced margin for WAV, leading to higher WFP. Variation-tolerant PSA topology is able to unlock the key limit factors of cross-point RRAM arrays, and therefore, may enable higher net capacity even after considering the non-programmable (unused) cells. Fig. 10(d) shows the allowed net capacity under conventional full-size and PSA topologies. $V_{DD}$ is fixed as 4.0 V and the array sizes are screened to hit the target 1% WFP for each topology. The net capacity for PSA topology includes the capacity loss due to non-programmable cells, yet a higher net capacity is obtained over conventional design due to the significantly improved variation tolerance.

Optimization of RRAM-based memory systems relies on the understanding of device and circuit interaction with an emphasis on variations. Fig. 11 summarizes the cross-layer design efforts discussed throughout this paper, which can also serve as guidelines for variation-aware RRAM optimization. Owing to a collaborative effect of device properties, circuit topologies, and operation conditions, device-level variations are nonlinearly 'translated' into circuit-level stochastic behaviors. These circuit behaviors are characterized by the statistics related to functionality, reliability, and energy. A full understanding of how device variability would affect these metrics will provide insights to guide device engineering. Holistically, the optimization space lies in the knowledge-based device tuning, architecture optimization, and tradeoffs between different metrics by circuit design knobs.

## V. Conclusion

Using a MC compact model of the RRAM, the device and circuit interaction analysis reveals the trend, dependency, and main contributors of stochastic behaviors in cross-point RRAM arrays. Scaling-up of array size, scaling-down of interconnect and $V_{DD}$, choice of bias schemes and data sparsity, play dominant roles in array functionality, reliability, and energy consumption. V/3 scheme has lower WFP, but is also more sensitive to increased device variations. V/2 scheme is more energy efficient from a statistical perspective. More generally, for V/*n* bias schemes, a larger *n* gives a lower WFP at the cost of worse energy efficiency and sensitivity to device variations. Design implications inspired by the circuit behaviors and metrics tradeoffs are discussed throughout the paper and exemplified in the case study of array topology optimization. This work provides a new optimization methodology for RRAM memory systems that require awareness of variations across all design layers.

## Acknowledgement

The authors would like to thank Z. Fang from IME, A*STAR, Singapore, H.-Y. Chen from Stanford University, and Z. Chen from Peking University, China, for providing RRAM samples used in this work.

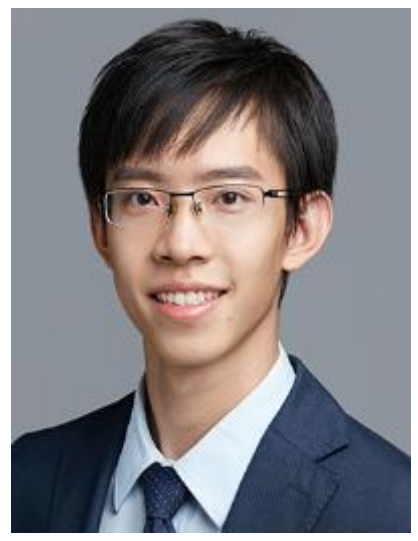

**Haitong Li** (S'14) received B.S. degree in Microelectronics from Peking University, Beijing, China, in 2015, and M.S. in Electrical Engineering from Stanford University, USA, in 2017.

He is currently pursuing the Ph.D. degree with the Department of Electrical Engineering, Stanford University, USA. His research interests include in-memory computing and neuromorphic computing with resistive RAM (RRAM).

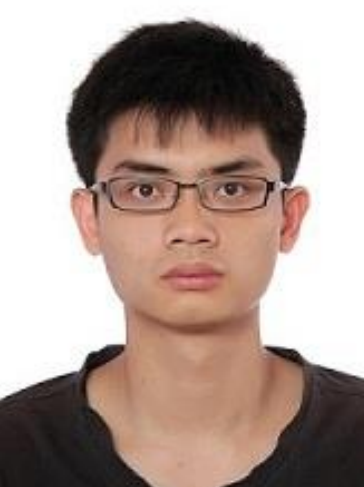

**Peng Huang** (S'10) received the B.S. degree from Xidian University, Xi'an, China, in 2010, and the Ph.D. degree from Peking University, Beijing, China, in 2015.

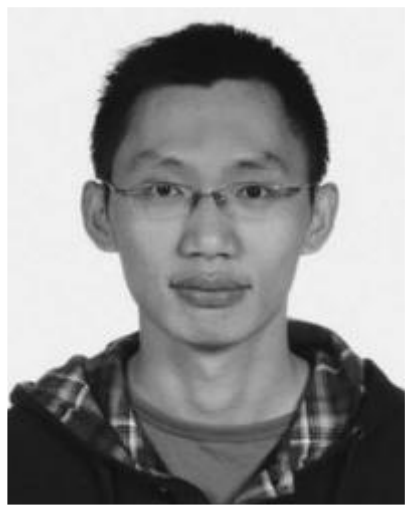

**Bin Gao** (S'08–M'14) received the B.S. degree in physics in 2008 from Peking University, Beijing, China, and received the Ph.D. degree in microelectronics from Peking University in 2013.

He is currently an assistant professor with the Department of Electrical Engineering, Tsinghua University, Beijing, China.

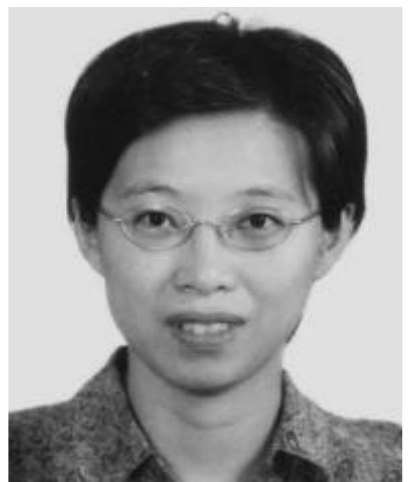

**Xiaoyan Liu** received the Ph.D. degree in microelectronics from Peking University, Beijing, China, in 2001.

She is currently a Professor with the Institute of Microelectronics, Peking University, Beijing, China.

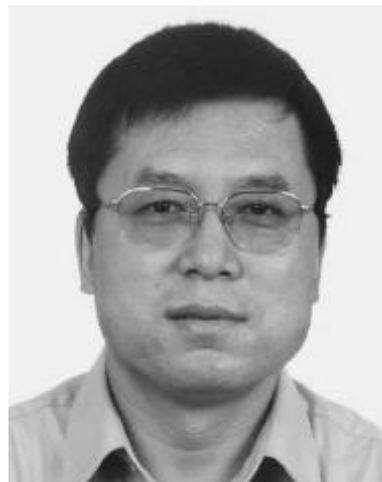

**Jinfeng Kang** received the Ph.D. degree in solid-state electronics from Peking University, Beijing, China, in 1995.

He is currently a Professor with the Institute of Microelectronics, Peking University, Beijing, China.

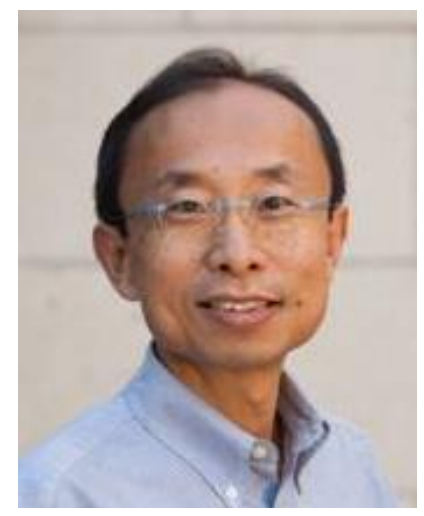

**H.-S. Philip Wong** (F'11) received the B.Sc. (Hons.) degree from The University of Hong Kong, Hong Kong, the M.S. degree from Stony Brook University, Stony Brook, NY, USA, and the Ph.D. degree from Lehigh University, Bethlehem, PA, USA.

He joined Stanford University, Stanford, CA, USA, in 2004, as a Professor of Electrical Engineering, where he is currently the Willard R. and Inez Kerr Bell Professor with the School of Engineering.